\documentstyle[pra,aps]{revtex}  \begin{document}
\draft
\title{\bf Quantum counter erasure}
\author{Fedor Herbut\cite{email} }
\address{Faculty of Physics, University of
 Belgrade,\\ 11001 BEOGRAD, P.O. Box 368, Yugoslavia\\ and the Serbian
Academy of Sciences and Arts\\ Beograd, Knez Mihajlova 35}

\maketitle
\begin{abstract}  Interference comes from coherent mixing. It can be
suppressed by entanglement, and the latter can be erased so as to revive
interference. If the entanglement is a mimal-term one (with minimal-term
mixing), as is the case in most thought and real experiments reported,
there appears the possibility of counter erasure and counter interference.
This peculiar phenomenon of minimal-term mixing and minimal-term
entanglement is investigated in detail. In particular, all two-term
mixings of an (arbitrary) given minimal-term mixed state are explicitly
exhibited. And so are their possible laboratory realizations in terms of
distant ensemble decomposition.\\
\end{abstract}

\section{INTRODUCTION}

	 In order to gain concrete experimental notions, we start by
discussing the well-known \it two-slit interference experiment \rm [1],
which is theoretically the simplest and best known example of interference.

Let the indices 1 and 2 refer to the two slits, and let $|\psi_1\rangle$
and $|\psi_2\rangle$ be the {\it spatial state vectors} of the photon having
traversed only the first or only the second slit respectively. Then the
superposition (also called coherent mixture) of these two state vectors,
i. e.,
$$ | \psi \rangle \equiv (1/2)^{1/2} \Big(| \psi_1 \rangle
+ | \psi_2 \rangle \Big) \eqno{(1)}$$
\noindent
is the \it interference state vector \rm corresponding to both slits
being open. (The term "interference" actually refers to the interference
pattern on the detection screen.) \\
\indent
To bring in entanglement, we assume that the photons pass a
horizontal linear polarizer at slit 1 and a vertical one at slit 2 [2].
The entangled two-subsystem (but one-photon) state vector is then,
in obvious notation:
$$ |\chi \rangle \equiv \Big(|H \rangle | \psi_1 \rangle
+ |V \rangle | \psi_2 \rangle \Big). \eqno{(2)} $$
\indent
We are dealing with a minimal-term entanglement (two terms
only). The state of the subsystem of spatial degrees of freedom is now
an improper mixture [3]
$$\rho_s\equiv Tr_p|\chi \rangle \langle \chi |=(1/2)\Big( \mid \psi_1\rangle \langle \psi_1|+\mid \psi_2\rangle \langle \psi_2|\Big) \eqno{(3)} $$
\noindent
as easily seen. The symbol  $ \rho_s$ denotes the state operator
(reduced statistical operator) of the spatial subsystem, and $"Tr_p"$
denotes the partial trace over the (linear) polarization degree of
freedom of the photon. Also the mixture in (3) is a minimal-term one.

The entanglement in (2) suppresses the interference replacing the
interference state (1) by the nonintereference one given by (3).

The entanglement (2) contains the so-called "which-path" memory,
because, in principle, measuring only if the linear polarization is
horizontal or vertical, one reestablishes $| \psi_1 \rangle$ or
$| \psi_2 \rangle$ respectively. For example, if the polarization
turns out to be horizontal, then, according to the so-called
L\"{u}ders formula for ideal measurement [4], [5], one has the
following disentanglement:
$$ | \chi \rangle \quad \rightarrow \quad c\Big( \mid H \rangle \langle H| \otimes 1 \Big) \mid \chi \rangle = |H \rangle |\psi_1 \rangle , \eqno{(4)} $$
\noindent
where c is a normalization constant.

Before this "which-path" measurement is performed, there is a
(potential) complementarity in $| \chi \rangle$ because it provides
also a complementary memory, on ground of which one can revive the
suppressed interference in the distant subsystem [6], [7].

This revival is possible because, as easily checked, one can rewrite
the \it same \rm composite-system state vector given by (2) as follows:
$$ |\chi \rangle =(1/2)^{1/2} \Big( |45^0 \rangle | \psi \rangle +
\mid -45^0 \rangle |\psi^c \rangle \Big), \eqno{(5)} $$
\noindent
where, e.g., $|45^0 \rangle$ is the polarization state at $45^0$
between horizontal and vertical, $| \psi \rangle$ is given by (1),
and, what we call the \it counter-interference state \rm (for reasons
seen below), $| \psi^c \rangle$ is defined by
$$ |\psi^c\rangle \equiv (1/2)^{1/2}\Big(|\psi_1 \rangle -|\psi_2\rangle \Big) . \eqno{(6)} $$
\noindent
Further, the corresponding linear polarization state vectors at the
given angles, evidently, satisfy
$$ |45^0 \rangle = (1/2)^{1/2}\Big(|H \rangle + |V \rangle \Big), \quad
\mid -45^0 \rangle = (1/2)^{1/2}\Big( |H\rangle -|V\rangle \Big) . \eqno{(7a,b)} $$
\indent
If one measures the linear polarization at $45^0$ or at $-45^0$ (since
$ \langle -45^0|45^0 \rangle =0,$ this is, essentially, an observable),
and if the former result is obtained, then, on account of (5), the
following disentanglement takes place (cf. [4] or [5]):
$$ |\chi \rangle \quad \rightarrow \quad \mid 45^0 \rangle |\psi \rangle .$$

Thus, the spatial interference state $| \psi \rangle$ is revived.
This phenomenon is called \it quantum erasure \rm [8], because the
"which-path memory" in the entanglement in $| \chi \rangle $,
which suppresses the interference, is erased.

\indent
If in the $45^0$-angle linear polarization measurement the result is
$-45^0$, then the L\"{u}ders formula gives $ | \psi^c \rangle $,
i. e., it is the counter-interference state that is revived.

\indent
As to what is actually observed in the laboratory, one cannot "see"
the interference state $| \psi \rangle $  itself (cf. (1)) in full.
One usually observes an interference pattern implied by
$ | \psi \rangle $ (on a detection screen). The pattern is actually
the localization probability distribution:
$$p_i(\mbox{\bf r})\equiv |\psi (\mbox{\bf r})|^2=(1/2)\Big( \mid \psi_1(\mbox{\bf r})|^2+|\psi_2(\mbox{\bf r})|^2+\psi_1^*(\mbox{\bf r})\psi_2(\mbox{\bf r})+\psi_1(\mbox{\bf r})\psi_2^*(\mbox{\bf r})\Big) , \eqno{(8a)} $$
\noindent
where "i" refers to interference, and $ \psi (\mbox{\bf r})
\equiv \langle \mbox{\bf r}| \psi \rangle $ is determined by (1), etc..\\
\indent
It follows from (5) that the state $\rho_s $ of the spatial subsystem (cf (3)) can also be written as
$$ \rho_s=(1/2)\Big( |\psi \rangle \langle \psi |+\mid \psi^c \rangle \langle \psi^c| \Big) . \eqno{(9)} $$
\indent
The probability distribution defined by the counter-interference state
$| \psi^c \rangle$, what we call \it counter interference \rm,
and the one defined by the incoherent mixture $ \rho_s$ are respectively:
$$ p_i^c( \mbox{\bf r}) \equiv | \psi^c( \mbox{\bf r})|^2 =$$
$$(1/2) \Big( |\psi_1( \mbox{\bf r})|^2+\mid \psi_2 ( \mbox{\bf r})|^2 - \psi_1^*( \mbox{\bf r})\psi_2( \mbox{\bf r})-\psi_1( \mbox{\bf r})\psi_2^*( \mbox{\bf r}) \Big) , \eqno{(8b)} $$
$$ p( \mbox{\bf r}) \equiv \langle \mbox{\bf r}| \rho_s | \mbox{\bf r} \rangle =
(1/2)\Big( p_i( \mbox{\bf r})+p_i^c( \mbox{\bf r}) \Big) =(1/2) \Big( | \psi_1( \mbox{\bf r})|^2+|\psi_2( \mbox{\bf r})|^2 \Big) . \eqno{(8c)} $$
\indent
The empirical, i. e., \it ensemble view \rm of the phenomena of
interference and counter-interference consists in realizing that, on
account of the measurement of $\Big(|45^0 \rangle \langle 45^0
| \otimes 1 \Big)$ on \it each \rm individual photon of a laboratory
ensemble that is described by $| \chi \rangle$ given by (2), the (improper [3])
ensemble of spatial subsystems described by $\rho_s$ breaks up into \it
two subensembles \rm (cf. the first sum in (8c)), each of which causes
interference on the detection screen (cf. (8a) and (8b) respectively),
but which are counter cases of each other in the sense that the two
interferences  cancel (cf (8c)). \\
\indent
A thought experiment in which the above mentioned linear polarizers
at the slits are replaced by maser cavities was given by Scully et al. in
[8]. The authors actually introduce \it quantum erasure \rm in a
pioneering way explaining the revival of the interference state
$| \psi \rangle$. With a slight modification of the experiment one can
revive the counter-interference state $| \psi^c \rangle$ instead of
$| \psi \rangle$. The first real experiment of quantum erasure was
attempted in [9]. It turned out [10] that it was erasure in a
somewhat broader sense.\\
\indent
Actually, the entangled composite-system state $| \chi \rangle$
contains \it a nondenumerable infinity of spatial states that can,
in principle, be revived \rm . The rivival takes place via the
measurement of an opposite-subsystem observable. Since the entanglement
is a minimal-term one, this observable is a yes-no measurement, and
the revived states appear \it in pairs, the counter states of each other
\rm . The "which-way" states $| \psi_1 \rangle$ and $| \psi_2 \rangle$
on the one hand and the interference state $| \psi \rangle$ and the
counter-interference state $| \psi^c \rangle$ on the other are examples of
counter states of each other.\\
\indent
We explore this phenomenon in detail in this study.

\section{COUNTER STATES IN MINIMAL-TERM MIXTURES}

In this section the following question is given an answer: \\
\indent \it
How to classify, i. e., enumerate (in a bijective way) explicitly the set
of all mathematically possible decompositions of a given minimal-term
mixture (like $\rho_s$ in (3)) into two pure states? \rm \\
\indent
This question is studied with a view to find out (in the next section) how
one can revive any of the two pure states of any of the mentioned
decompositions by a yes-no measurement on the opposite subsystem. \\
\indent
Let $\rho$ be a given minimal-term mixture state operator, i. e., one that
can be written in the \it spectral form \rm :
$$ \rho = r|1 \rangle \langle 1| + (1-r)|2 \rangle \langle 2|, \eqno{(10)} $$
\noindent
where $0<r\leq (1/2)$. It is known [11] that each state vector from the
range of $\rho$, and only such state vectors, can appear in a decomposition
of $\rho$. We want to find out about the counter state vectors and the
corresponding statistical weights. \\

Our \it answer \rm to the above question goes as follows:\\
\indent
Let (10) be given. Let, further,
$$|\phi \rangle \equiv p|1 \rangle +(1-p^2)^{1/2}e^{i\vartheta}|2\rangle, \eqno{(11a)} $$
\noindent
with any values from the intervals
$$ 0 \leq p \leq 1, \qquad 0 \leq \vartheta <2 \pi, \eqno{(11b)} $$
\noindent
be an (up to a phase factor) arbitrary state vector from the range of
$\rho$. Then \it there exists one and only one decomposition of $\rho$
into two pure states in which $| \phi \rangle \langle \phi|$ appears. \rm
It is
$$ \rho = w| \phi \rangle \langle \phi |+
(1-w) | \phi^c \rangle \langle \phi^c | \eqno{(12)} $$
\noindent
where
$$ w \equiv r(1-r) \Big/ \Big( p^2(1-r)+(1-p^2)r \Big), \eqno{(13)} $$
\noindent
and
$$ | \phi^c \rangle \equiv \Big[ (r-wp^2) \Big/ (1-w) \Big]^{1/2}|1 \rangle \enspace + $$
$$ \Big[ \Big( (1-r)-w(1-p^2) \Big)
 \Big/ (1-w) \Big]^{1/2}e^{i(\vartheta+ \pi )}|2 \rangle . \eqno{(14)} $$

\indent
The claims made are shown to follow as an immediate consequence of a
wider lemma stated and proved in Appendix 1. \\
\indent
To be practical, we shall call decomposition (12) of $\rho$ in the context of (10)-(14)
\it "the p,$\vartheta $-decomposition". \rm Thus, all decompositions of
a given minimal-term mixture $\rho$ ca be classified or enumerated by
the two parameters p and $ \vartheta $. \\
\indent
The counter state $| \phi^c \rangle $ and the statistical weight
$w$ are uniquely implied by the state operator $ \rho $ and $| \phi \rangle $.
The state vectors $| \phi \rangle $ and $| \phi^c \rangle $ are counter states of each other, i. e., if one is written as (11a), then the other takes the form (14). \\
\indent
Further, a more detailed examination of the answer given reveals the
following \it peculiarities in the above
relations: \rm \\
\indent
(i) If the characteristic value $r$ of $\rho $ is nondegenerate, or,
equivalently, if $r<(1/2)$, then relation (13) establishes a
monotonously decreasing bijection of the interval [0,1] of the values of
$p$ onto the interval $[r,(1-r)]$ of the values of $w$. (Namely,
$dw/d(p^2)<0$.)  \\
\indent
(ii) If $r=(1-r)=1/2$, then $p$ and $ \vartheta $ can still take all
values from their respective intervals (11b), but always $w=1-w=1/2$.
In this case the counter state takes the simple form
$$ | \phi^c \rangle = (1-p^2)^{1/2}|1 \rangle - pe^{i \vartheta }|2 \rangle
\eqno{(15)} $$
\noindent
and $| \phi^c \rangle $ is \it orthogonal to \rm $| \phi \rangle $.
Decomposition (12) is now a spectral form of $ \rho $ (just like (10)).
In this case, \it every decomposition \rm of $ \rho $ into pure states is an \it orthogonal \rm one (a spectral form), and \it there are no other decompositions into pure states \rm . Further, \it every \rm orthogonal decomposition of the range $R( \rho )$ gives also a decomposition of $\rho$, and vice versa. \\
\indent
(iii) Always $r\leq w\leq (1-r)$. The equality r=w is observed if
and only if $p=1$, then $|\phi \rangle =|1\rangle$; whereas $w=(1-r)$
if and only if $p=0$, and then $|\phi \rangle =|2\rangle $. (These are consequences of (i) and (11a).)\\
\indent
In case of nondegenerate $r$, peculiarity (iii) implies that the spectral form (10) is the mixture in which the most dominant pure state (i. e., the one with the largest statistical weight) and the least dominant one are exhibited. All other mixture forms (i. e., $p, \vartheta$-decompositions) of the given state operator $\rho$ are less extreme. \\
\indent
 In case of nondegenerate $r$, it, further, ensues from peculiarity (i) that for any
\it a priori \rm given $w \in (r,1-r)$, there exists a family of
$p, \vartheta$-decompositions that give this $w$ value: The (unique) value of
$p$ is obtained by solving (13) for $p$, and $ \vartheta $ is arbitrary. In particular,  $w=1/2$ is obtained with $p=r$. \\
\indent
Before we tackle (in the next section) the problem of how to perform
empirically decomposition
(12) of an empirically given (subsystem) state $ \rho $, it should be noted that
this decomposition may find application in various problems. For instance,
$\rho$ may be the state operator of a composite system, and
$| \phi \rangle$ (cf (11a)) an uncorrelated state vector. The evaluation
of the counter state  $| \phi^c \rangle$ \ (cf (14)) is then of interest
because it decomposes $\rho$ into a separable and an inseperable state
(cf [12] and [13]).

\section{WHICH YES-NO MEASUREMENT GIVES RISE TO A GIVEN DECOMPOSITION? }

  The state vectors $| \phi \rangle$ and $| \phi^c \rangle$ in
decomposition (12) are in general not orthogonal. Hence, one cannot
produce decomposition (12) by measurement in the laboratory because
this always ends up in orthogonal states. \\
\indent
  Nevertheless, these decompositions \it do have physical meaning
\rm in terms of so-called \it distant state decomposition \rm (empirically
distant ensemble decomposition): \\
\indent
  One views the system on hand as a subsystem of a two-subsystem composite
system, and one envisages the state vector $| \omega \rangle$ of the latter that implies
the \it a priori \rm given state operator $\rho$ (cf (10)) as its subsystem
state operator $\rho =Tr_o| \omega \rangle \langle \omega|$ (the letter
"o" in the index of the partial trace applies to the "opposite" subsystem).
Then, arguing along the lines presented in the Introduction for the Young
two-slit interference, an opposite-subsystem yes-no measurement on the
composite system in the state $| \omega \rangle$ may leave the subsystem
state $\rho$ decomposed precisely as given in the
$p, \vartheta$-decomposition (12). This is what we investigate in detail in
this section. \\
\indent
 We have the given minimal-term mixture $\rho$ in spectral form (10). We
write $| \omega \rangle$ expanded in the characteristic basis
$\Big\{ |1 \rangle ,|2 \rangle \Big\}$ of $\rho$ with positive
expansion coefficients:
$$|\omega \rangle =r^{1/2}|1\rangle_o|1\rangle +(1-r)^{1/2}|2\rangle_o|2\rangle . \eqno{(16)} $$
\indent
  This is a so-called Schmidt biorthogonal expansion (cf section 4 in [14]
or see [15]). The vectors $\Big\{|1 \rangle_o,|2 \rangle_o \Big\}$ are
orthogonal state vectors in the state space of the opposite subsystem. (One may define $|\omega \rangle$ via (16) by choosing any such
subbasis.)\\
\indent
  A suitable observable on the opposite subsystem that is a yes-no
one on $|\omega \rangle$ has the following
spectral form:
$$A_o=a_1|\mu_1\rangle_o\langle \mu_1|_o+a_2|\mu_2\rangle_o\langle \mu_2|_o,\quad a_1\not= a_2, \eqno{(17)} $$
\noindent
where (the state vectors) $| \mu_1 \rangle_o$ and $| \mu_2 \rangle_o$ are required to be (mutually orthogonal) linear combinations of
$|1 \rangle_o$ and $|2 \rangle_o$. \\
\indent
  One should note that one of the exhibited characteristic values
of $A_o$ can be zero if
the opposite-subsystem state space is two dimensional. But if it is three-
or more dimensional, then both $a_1$ and $a_2$ must be nonzero. Then, as
evident from (17), $A_o$ necessarily has zero in its spectrum (though
it is not exhibited in (17)). \\
\indent
  We treat the characteristic values $\Big\{ a_1,a_2 \Big\}$ as irrelevant,
i. e., we consider the whole class of observables having the same characteristic vectors $\Big\{| \mu_1 \rangle_o,| \mu_2 \rangle_o \Big\}$  as (essentially) one
observable (as it is often done). \\
\indent
  Further, the characteristic vectors can be written in the following
suitable form:
$$|\mu_1\rangle_o =q|1\rangle_o+\Big( 1-q^2\Big)^{1/2} e^{i\lambda}\mid 2 \rangle_o, \eqno{(18a)} $$
$$ 0 \leq q \leq 1, \qquad 0 \leq \lambda <2 \pi; \eqno{(18b)} $$
$$ | \mu_2 \rangle_o = \Big( 1-q^2 \Big)^{1/2}|1 \rangle_o-q e^{i \lambda }|2\rangle_o; \eqno{(19)} $$
where $\{|1\rangle_o,|2\rangle_o\}$ are determined by (or determine)
the composite-system state vector $|\omega \rangle$ (cf (16)).\\
\indent
 We call the $\Big( A_o \otimes 1 \Big)$ measurement on the composite
system in the state $| \omega \rangle$ \it the $q,\lambda$-measurement. \\ \rm

  Now, one can make the following \it claim \rm , which {\it answers}
the question from the title of the section:

 If a $p, \vartheta$-decomposition (12) of a given minimal-term
mixture state operator $\rho$ (cf (10)) is given and a minimal-term
entanglement composite system state vector $| \omega \rangle$ (cf (16))
implying $\rho$ as its subsystem state operator is also given, then
\it the following $q, \lambda$-measurement performed on $| \omega \rangle$, and no other one, gives rise to the mentioned $p,\vartheta$-decomposition: \rm
$$ q = \Big( w\Big/ r\Big)^{1/2} p .\eqno{(20a)} $$
\indent
where $w$ is the statistical weight of $|\phi \rangle \langle \phi|$ in
decomposition (12) given by (13), and
$$ \lambda = 2\pi -\vartheta . \eqno{(20b)} $$
\indent
 The claim is proved in Appendix 2. \\

 Inverting the question from the title of the section, the (second) {\it answer} is as
follows: \\
\indent \it
 A given $q,\lambda$-measurement \rm (cf (17)-(19)) on the composite-system
state $|\omega \rangle$ given by (16) \it gives rise to the following
$p,\vartheta$-decomposition \rm (12) of the subsystem state operator
$\rho$ implied by $|\omega \rangle$:
$$ p = \Big( r\Big/ w\Big)^{1/2}q , \eqno{(21a)} $$
\noindent
where $w$ is the statistical weight given by (13), which can now more
suitably be written as
$$ w = (2r-1)q^2 + (1-r) ; \eqno{(21b)} $$
\noindent
and, finally,
$$ \vartheta = 2\pi -\lambda . \eqno{(21c)} $$
\indent
 The validity of this claim is proved in Appendix 3.\\

 The two claims made establish a \it correspondence \rm between
the set of all decompositions (12) and the set of all suitable yes-no
measurements (cf (17)) on the opposite subsystem. "Suitability" here
means that the two characteristic state vectors $|\mu_1 \rangle_o$ and
$|\mu_2 \rangle_o$ exhibited in (17) span the range of the
opposite-subsystem state operator of $|\omega \rangle$
(cf (16)), and, as a consequence, one can expand $|\omega \rangle$
in them (cf (22) below).

\section{CONCLUDING REMARKS}
 Let us return from detail to the global conceptual view. \\
\indent
 If a composite-system state vector $|\omega \rangle$ is given
in a two-term Schmidt biorthogonal expansion (16), we can expand it in
any orthonormal basis $\Big\{|\mu_1\rangle_o,|\mu_2 \rangle_o \Big\}$ in the
range of the opposite-subsystem state operator $\rho_o\quad \Big(\equiv
Tr|\omega \rangle \langle \omega |\Big)$ ("$Tr$" denotes here the partial trace
over the subsystem at issue):
$$ |\omega \rangle = |\mu_1 \rangle_o|\phi_1'\rangle+|\mu_2 \rangle_o|\phi_2'\rangle . \eqno{(22)} $$
\noindent
(The vectors $|\phi_i' \rangle$ , i=1,2 , are, in general, not normalized,
i. e., they are not state vectors). Then the nonselective (or all-results)
version of ideal measurement of the observable $\Big( A_o \otimes 1\Big)$,
where $A_o$ is given by (17) in terms of the basis considered, converts
$|\omega \rangle$ into the mixed state
$$ \langle \phi_1'|\phi_1' \rangle |\mu_1 \rangle_o\langle \mu_1|_o
\otimes \Big( |\phi_1'\rangle \langle \phi_1'|\Big/ \langle \phi_1'|
\phi_1'\rangle \Big) + $$
$$\langle \phi_2'|\phi_2' \rangle |\mu_2\rangle_o
\langle \mu_2|_o \otimes \Big( |\phi_2'\rangle \langle \phi_2'|\Big/
\langle \phi_2'|\phi_2' \rangle \Big)  \eqno{(23)} $$
\noindent
with $\langle \phi_i'|\phi_i'\rangle$, i=1,2, as the statistical weights
(cf the L\"{u}ders formula (4) that applies to selective or
particular-result measurement).This composite-system mixture implies the
same subsystem state $\rho$ as $|\omega \rangle$ does (as one can see
from (22) and (23)), and it also implies its decomposition into pure states:
$$ \rho = \langle \phi_1'|\phi_1'\rangle
\Big( |\phi_1'\rangle \langle \phi_1'|\Big/ \langle \phi_1'|\phi_1'
\rangle \Big) + \langle \phi_2' |\phi_2' \rangle \Big( |\phi_2'\rangle \langle \phi_2'|\Big/ \langle \phi_2'|\phi_2'
\rangle \Big) . \eqno{(24)}  $$
\indent
In the two answers in the preceding section we have
$$ \langle \phi_1'|\phi_1'\rangle = w, \qquad \langle \phi_2'|\phi_2'
 \rangle = 1-w ; \eqno{(25a,b)} $$
$$ |\phi_1'\rangle \Big/ \Big( \langle \phi_1'|\phi_1'\rangle^{1/2}\Big)
= |\phi \rangle ,\qquad |\phi_2'\rangle \Big/ \Big( \langle \phi_2'|\phi_2'
\rangle^{1/2}\Big) = |\phi^c\rangle . \eqno{(25c,d)} $$
\indent
 The state decompositions (23) and (24) are \it actual \rm (not just
potential or mathematically possible like, e. g., the expansion (22))
because if one takes into account the (suppressed) states of the
measuring instrument that has performed the $\Big( A_o\otimes 1\Big)$
-measurement, different "positions" of the "pointer" (symbolically stated)
correspond to the two terms. \\
\indent
Finally, let us discuss the special case when (24) is an \it orthogonal
decomposition \rm of $\rho$, hence, in principle, a measurement. It is
called \it distant measurement \rm [15], [16], because the subsystem is not dynamically
influenced by the opposite-subsystem measurement.\\
\indent
If the characteristic value $r$ of $\rho$ is \it not degenerate \rm, (10) is the only orthogonal
decomposition of $\rho$. In this case distant measurement takes place
if and only if $|\mu_i\rangle_o =|i\rangle_o$,
$i=1,2$ (cf (16)), and we are dealing with a common characteristic subbasis of $A_o$ and $\rho_o$.

Commutation of $A_o$ with $\rho_o$ is a necessary and
sufficient condition for distant measurement for a general entangled
two-subsystem state vector as proved in [15] and [16].\\
\indent
If $r$ is \it degenerate \rm, every choice of $A_o$ (as long as $|\mu_1\rangle_o$ given by (18a) and $|\mu_2\rangle_o$ given by (19)
span the range of $\rho_o$) leads to distant measurement because the
state operator is a constant in $R\left(\rho_o\right)$, and,
hence, $A_o$ always commutes with it.\\
\indent
A beautiful realization of essentially the entangled composite
state vector $|\chi \rangle$ given by (2) in a \it real experiment \rm
has been reported [17]:\\
\indent
Instead of two slits, there are two processes of parametric down
conversion. We'll disregard, say, the so-called signal out of the pair of
down-converted photons, and speak only about the so-called idler. The
idler from the first process is reflected back so that it may spatially
overlap with the idler created in the second process and thus approach a
detector. Writing the state vector of the former as $|\psi_1\rangle$,
and that of the latter as $|\psi_2\rangle$, the photon may stem either
from the first or from the second process, and thus one obtains the
above interference state $|\psi \rangle$ given by (1). The phenomenon of
interference is observed by moving the mentioned reflecting mirror,
and thus changing $|\psi_1\rangle$ and changing the detection probability. \\
\indent
 Both signal and idler are vertically polarized in the very processes
of down-conversion. The role of the (mutually orthogonal)
polarizers at the slits (see the Introduction) is here played by a quarter-wave plate that is
put in the way of the idler from the first process (to be traversed to
the mirror and back). It serves to rotate the polarization from vertical
to horizontal. Thus, essentially the above entangled state $|\chi \rangle$
(cf. (2)) comes about. \\
\indent
 Putting an analyzer at $45^0$ in front of the detector, erasure is
observed on the photons that pass the analyzer and reach the detector
(cf. (5)). If the analyzer is at $-45^0$, then the counter-interference
state $|\psi^c\rangle$ is obtained out of $|\chi \rangle$. Other angles
of the analyzer would, if the photon passes, give rise to, or \it
distantly prepare, \rm the spatial state in other linear
combinations of $|\psi_1\rangle$ and $|\psi_2\rangle$. \\
\indent
 And all this is only a small part of the mentioned experiment
[17]. Incidentally, it may be compared, at least partially, with
a previous experiment [18], because they both give realization  to
Franson's idea [19] of superposing (coherently mixing), essentially,
different instants of creation of the photon, which comes about due to
some spatial detour that exceeds the coherence length. But in the recent
experiment [17] polarization is included and manipulated in a practical
way, and thus Ryff's idea [20] of observing quantum erasure in Franson's
experiment can be considered realized.\\
\indent
As a matter of fact, the experiment [17]
seems to be independent of these ideas, because the corresponding
articles are not among the references of [17].
\appendix
\section{}

  \rm
 We rewrite the relations (10), (11a), (13) and (14) in a redundant,
but more compact and for proof more suitable form:
$$ \rho = r |1\rangle \langle 1| + r' |2\rangle \langle 2| , \quad
r' = 1-r; \eqno{(A.1)} $$
$$ |\phi \rangle \equiv p |1\rangle + p'e^{i\vartheta} |2\rangle , \quad
p' \equiv \Big( 1-p^2 \Big)^{1/2} ; \eqno{(A.2)} $$
$$ w \equiv rr' \Big/ \Big( p^2r'+p'^2r\Big) , \quad w'\equiv 1-w;
\eqno{(A.3)} $$
$$ |\phi^c\rangle \equiv \Big[ \Big( r-wp^2\Big) \Big/ w'\Big]^{1/2}
|1\rangle + \Big[ \Big( r'-wp'^2\Big) \Big/ w'\Big]^{1/2}e^
{i(\vartheta+\pi )} |2\rangle . \eqno{(A.4)} $$
\indent \it
Lemma A1. \rm Let a parameter $s$ be given such that $0<s\leq 1$. Then for
each value of $s$ from the given interval, one can decompose $\rho$
uniquely as follows:
$$ \rho=ws|\phi \rangle \langle \phi |+(1-ws)\rho ', \eqno{(A.5)} $$
$$ \rho '\equiv \Big( (w-ws)\Big/ (1-ws)\Big) |\phi \rangle \langle
\phi | + \Big( (1-w)\Big/ (1-ws)\Big) |\phi^c\rangle \langle \phi^c
|. \eqno{(A.6)} $$
\indent
 If $s>1$, then there exists no statistical operator $\rho '$ such
that decomposition (A.5) is valid.\\
\indent \it
 Proof. \rm Replacing (A.6) in (A.5), the latter reduces to (12):
$$ \rho =w|\phi \rangle \langle \phi |+w'|\phi^c\rangle \langle \phi^c|.
 \eqno{(A.7)} $$
\indent
 Evidently, (A.5) is valid if and only if so is (A.7). Checking this
relation, one easily obtains
$$ \langle 1|LHS|1\rangle =\langle 1|RHS|1\rangle $$
\noindent
and
$$ \langle 2|LHS|2\rangle =\langle 2|RHS|2\rangle . $$
\noindent
Further, $\langle 1|LHS|2\rangle =0$, and
$$ \langle 1|RHS|2\rangle =wpp'e^{-i\vartheta}-\Big( r-wp^2\Big)^{1/2}
\Big( r'-wp'^2\Big)^{1/2}e^{-i\vartheta}= $$
$$wpp'e^{-i\vartheta}-\Big( rr'-rwp'^2-r'wp^2+w^2p^2p'^2\Big)
^{1/2}e^{-i\vartheta}. $$
\noindent
Substituting here $rr'$ from (A.3), one obtains
$$ \langle 1|RHS|2\rangle =$$
$$wpp'e^{-i\vartheta}\enspace -\enspace \Big( r'wp^2 +rwp'^2-rwp'^2
-r'wp^2+w^2p^2p'^2\Big)^{1/2}e^{-i\vartheta}=0. $$
\indent
 The operator $\rho '$ is unique because it is determined by (A.5)
in terms of the rest of the entities in this relation. Assuming $s'>1$ and the validity of (A.5) with $s\equiv s'$ and $\rho '\equiv \rho ''$, where
$\rho  ''$ is some hypothetical statistical operator, we can write
(A.5) as follows:
$$ \rho =w|\phi \rangle \langle \phi |+(ws'-w)|\phi \rangle \langle \phi|
+(1-ws')\rho ''. $$
\noindent
Subtracting (A.7) from this, one obtains
$$ \Big( (ws'-w)\Big/ w'\Big) |\phi \rangle \langle \phi |+
\Big( (1-ws')\Big/ w'\Big) \rho ''=|\phi^c\rangle \langle \phi^c|. $$
\noindent
This is not possible due to the homogeneity of the state on the RHS and
the fact that $|\phi^c\rangle \not= |\phi \rangle$ (or else
$\rho =|\phi \rangle \langle \phi |$, which is not true because $\rho$ is
assumed to be a mixture). This {\it reductio ad absurdum} argument
proves that decomposition (A.5) with $s>1$ is not possible. \quad $\Box$

\it Corollary A1. \rm Decomposition (A.7) is the
only one that decomposes the mixture $\rho$ into two pure states one of
which is
$|\phi \rangle \langle \phi |$.\\
\indent \it
 Proof. \rm Let us assume \it ab contrario \rm that there exists
another decomposition
$$\rho =w'|\phi \rangle \langle \phi |+
(1-w')|\phi'' \rangle \langle \phi'' |. $$
\noindent
If $w'>w$, then we can rewrite this in the form of (A.5) with $s>1$, but, according to lemma A1, this is not possible. If $w'<1$, then we can, again,  put this
in the form of (A.5), but this time with $s<1$. Then, one, further,
obtains
$$ |\phi'' \rangle \langle \phi'' |=\rho '. $$
\noindent
This is not possible because $\rho'$ is a mixture (cf (A.6)).
Finally, if $w'=w$, then $|\phi'' \rangle \langle \phi'' |$ is determined
by the rest of the entities in the above decomposition. Thus, it
cannot differ from $|\phi' \rangle \langle \phi' |$ (cf (A7)).$\Box$

\section{}

Let a $p,\vartheta$-decomposition of $\rho$ (cf (10)-(14)) be given
together with a composite-system state vector $|\omega \rangle$ that
implies $\rho$ as its subsystem state operator (cf (16)). To evaluate
the corresponding yes-no measurement, we write (22) with (25a-d)
substituted in it:
$$ |\omega \rangle =w^{1/2}|\mu_1\rangle_o|\phi \rangle +
(1-w)^{1/2}|\mu_2\rangle_o|\phi^c\rangle . \eqno{(A.8)} $$
\noindent
Substituting here $|\omega \rangle$ from (16), partial scalar product
(see section 2 in [14] or see [15]) with $|\mu_1\rangle_o$ from the
left gives
$$ r^{1/2}\Big( \langle \mu_1|_o|1\rangle_o\Big) |1\rangle +
(1-r)^{1/2}\Big( \langle \mu_1|_o|2\rangle_o\Big) |2\rangle =
w^{1/2}|\phi \rangle $$
\noindent
on account of $\langle \mu_1|_o|\mu_2\rangle_o=0$. Inserting the explicit
forms of $|\phi \rangle$ and $|\mu_1\rangle_o$, i. e., (11a) and (18a)
respectively, one further has
$$ r^{1/2}q |1\rangle + (1-r)^{1/2}(1-q^2)^{1/2}e^{-i\lambda }|2 \rangle =
w^{1/2}p|1 \rangle + w^{1/2}(1-p^2)^{1/2}e^{i\vartheta}|2 \rangle $$
\noindent
or, putting the corresponding expansion coefficients on the two sides equal, one obtains
$$ r^{1/2}q = w^{1/2}p , \quad (1-r)^{1/2}(1-q^2)^{1/2}e^{-i\lambda}
= w^{1/2}(1-p^2)^{1/2}e^{i\vartheta} . \eqno{(A.9a,b)} $$
\indent
Relation (A.9a) can be rewritten as
$$ q = (w/r)^{1/2}p . \eqno{(A.10a)} $$
\indent
To evaluate $w$, we utilize relation (A.9b), where equality of the
norms, upon squaring, implies
$$ (1-r)(1-q^2) = w(1-p^2) . \eqno{(A.11)} $$
\noindent
Replacing here $q^2$ from (A.10a), one derives
$$ w = r(1-r)\Big/ \Big( p^2(1-r)+(1-p^2)r\Big) , \eqno{(A.10b)} $$
\noindent
which is, actually, relation (13). In relations (A.10a) and (A10b) the
dependence of $q$ on $p$ is expressed via $w$. \\
\indent
The phase factors in (A.9b) give the second part of our unique
solution:
$$ \lambda =2\pi -\vartheta . \eqno{(A.10c)} $$
\section{}
 Let a $q,\lambda$-measurement (cf (17)-(19)) be given together
with the composite-system state vector $|\omega \rangle$ determined by
(16) in which the measurement is to be performed. To evaluate the
corresponding $p,\vartheta$-decomposition of $\rho$, the state operator of
the second subsystem, we return to the argument presented in Appendix 2
leading to (A.9a) and (A.9b). These relations connect $p,\vartheta$ and
$q,\lambda$ independently of the fact which of them is given \it a priori.
\rm \\
\indent
Solving (A.10a) for $p$, we obtain
$$ p = (r/w)^{1/2}q , \eqno{(A.12a)} $$
\noindent
and solving (A.11) with (A.12a) for $w$, we end up with
$$ w = (2r-1)q^2 + (1-r) . \eqno{(A.12b)} $$
\indent
The second part of our unique solution comes from inverting
(A.10c):
$$\vartheta = 2\pi -\lambda . \eqno{(A.12c)} $$

\end{document}